\documentclass[%
reprint,
amsmath,amssymb,
aps,
]{revtex4-1}

\usepackage{graphicx}
\usepackage{dcolumn}
\usepackage{bm}

\usepackage[english]{babel}
\usepackage{color}
\begin{document}

\preprint{APS/123-QED}

\title{Supercontinuum generation assisted by dispersive waves trapping in dispersion-managed integrated silicon waveguides}
\author{Junxiong Wei$^{1,3}$}
\email{Junxiong.Wei@ulb.ac.be}
\author{Charles Ciret$^2$, Maximilien Billet$^{1,3}$, Fran\c{c}ois Leo$^1$, Bart Kuyken$^3$, Simon-Pierre Gorza$^1$}
\affiliation{ $^1$OPERA-Photonique CP 194/5, Universit\'e Libre de Bruxelles (ULB), Av.F.D. Roosvelt 50, B-1050 Bruxelles, Belgium \\
 $^2$Laboratoire de Photonique d’Angers EA 4464, Université d’Angers, 2 Bd. Lavoisier, 49000
Angers, France \\
$^3$ Photonics Research Group, Department of Information Technology, Ghent University-IMEC, B-9000, Ghent, Belgium}

\date{\today}

\begin{abstract}
Compact chip-scale comb sources are of significant interest for many practical applications. Here, we experimentally study the generation of supercontinuum (SC) in an axially varying integrated waveguide. We show that the local tuning of the dispersion enables the continuous blue shift of dispersive waves thanks to their trapping by the strongly compressed pump pulse. This mechanism provides new insight into supercontinuum generation
in dispersion varying integrated waveguides. Pumped close to 2.2 $\mu$m in the femtosecond regime and at a pulse energy of $\sim$ 4 pJ, the output spectrum extends from 1.1 $\mu$m up to 2.76 $\mu$m and show good coherence properties. Octave-spanning SC is also observed at an input energy as low as $\sim$ 0.9 pJ. We show that the supercontinuum is more robust against variations of the input pulse parameters and is also spectrally flatter in our numerically optimized waveguide than in fixed width waveguides. This research demonstrates the potential of dispersion varying waveguides for coherent SC generation, and paves the way for integrated low power applications, such as chip scale frequency comb generation, precision spectroscopy, optical frequency metrology, and wide-band wavelength division multiplexing in the near infrared.

\end{abstract}

\pacs{Valid PACS appear here}
\maketitle


\section{\label{sec:level1}Introduction}

Spectral broadening of light and generation of new frequency components are ultimate features of nonlinear optics. They have been intensively studied since the discovery of second harmonic generation in 1961 \cite{Franken_1961}. The particular nonlinear process known as supercontinuum (SC) generation occurs when narrowband incident pulses or continuous waves experience a dramatic spectral broadening \cite{Dudley_2010}. It was first reported by Alfano and Shapiro in 1970 \cite{Alfano_1970}. Since then it has been observed in a wide variety of nonlinear media, including solids \cite{Werner_2019}, liquids \cite{Bethge_2010}, gases \cite{Corkum_1987}, and various types of waveguides \cite{Kou_2018,Porcel_2017,Phillips_2011}. Today, broadband SC generation has found numerous applications such as high-precision spectroscopy \cite{Grassani_2019,Jahromi_2019,Eslami_2019}, frequency metrology and synthesis \cite{Carlson_2017,Lamb_2018,Singh_18}, ultra-short pulse generation \cite{Okamura_2015}, wavelength division multiplexing (WDM) communication \cite{Saghaei_2017}, or nonlinear optical spectroscopy and imaging in biophotonics \cite{Tu_2013,Bassi_2007,McConnell_2004}. The integration of SC sources on a chip, based on compact planar waveguide circuits rather than optical fibers, can provide compact, robust, and power efficient platforms for the aforementioned applications.

Both experimental and theoretical investigations of SC generation in waveguides have been a hot topic \cite{Kuyken_2011,Oh_2014, Liu_2016,Shams_2019, Singh_2015, Kuyken_2020}. In order to build integrated electronic and photonic devices on the same chip for mass market applications, optical waveguides should be fabricated together with silicon electronic components or be CMOS compatible. The silicon on insulator (SOI) platform, showing low propagation loss and high index contrast is one of the leading candidates to achieve this purpose. Silicon also has a large nonlinear coefficient, similar to that of chalcogenide glass \cite{Sanghera_2007} and about 200 times higher than silica \cite{Kim_1994}, facilitating ultra-compact and power-efficient nonlinear devices for SC generation. In 2007 Yin et al. \cite{Yin_2007} reported the first numerical simulations of SC generation in a 1.2 cm long crystalline silicon waveguide and predicted spectral broadening over 400 nm. In the same year, Hsieh et al. \cite{Hsieh_2007} published the first experimental demonstration. A spectral broadening of approximately 300 nm was obtained using 100 fs input pulses at 1.3 $\mu$m. Later, SC generation pumped at telecommunication wavelengths and spanning from 1200 to 1700 nm in a standard (220 nm high) SOI waveguide was reported \cite{Leo_2014}. In both cases, the two photon absorption (TPA) clamps the maximum power in the chip, thereby impeding the maximum achievable spectral broadening. However, the nonlinear loss can dramatically be reduced when the pump wavelength lies beyond $\sim$2200 nm \cite{Liang_2004,Jalali_2010,Bristow_2007}. The first demonstration of octave-spanning SC on a SOI chip resorted to input pulses near 2.5 $\mu$m and a 320 nm by 1210 nm cross section waveguide exhibiting anomalous dispersion at that wavelength \cite{Lau_2014}. In 2018 an octave-spanning SC in the near infrared, covering 1.124-2.4 $\mu$m, has been generated in SOI ridge waveguides \cite{Singh_2017}. In order to avoid the inherent nonlinear loss encountered in silicon in the near infrared and achieve broader SC spectra, the research moved toward other platforms such as silicon nitride \cite{Mayer_2015}, chalcogenide \cite{Xing_2018}, silicon-germanium \cite{Sinobad_2018}, indium gallium phosphide \cite{Dave_2015} and aluminium nitride \cite{Liu_2019}. In the context of SC generation, Guo et al. \cite{Guo_2018} reported a mid-infrared comb source from 2.5 $\mu$m to 4\,$\mu$m in silicon nitride (S$i_3$N$_4$) waveguides and, recently, a source of supercontinuum spanning from 400 nm to 2400 nm has been demonstrated in integrated lithium niobate waveguides \cite{Yu_2019}.

It is now well understood that the dynamics of on chip broadband SC generation pumped in the anomalous dispersion regime with ultrashort pulses is dominated by self-phase modulation and temporal pulse compression, followed by higher-order soliton fission and emission of resonant dispersive waves \cite{Yin_2007,Lin_2007}. In the absence of Raman induced self-frequency shift, the widespread strategy adopted to generate large spectra in these platforms merely consists in designing the cross-section of the waveguide so as for the dispersive waves to be emitted as far as possible from the pump wavelength. In the literature, these structures are referred as dispersion engineered waveguides. Inspired by previous works with optical fibers (see e.g. \cite{Hori_2004,Nathan_2005,Gorbach_2007,Kud_2008,Travers_2009}), cascaded waveguides, tapers and more complex dispersion managed integrated structures were recently investigated both numerically \cite{Hu_2015,Xiang_2016} and experimentally \cite{Ishizawa_2017,Singh_2019,Carlson_2019}. The motivation for studying dispersion varying waveguides is the ability to locally adapt the phase matching conditions required for the generation of new frequencies through four-wave mixing processes. This encompasses the generation of dispersive waves at different detunings from the pump wavelength, but also frequency shifting through collisions and trapping at event horizons. The second advantage is the possibility to ensure a high degree of coherence over the whole output spectrum by maintaining, at the beginning of the waveguide, a large anomalous dispersion or, equivalently, a low soliton order. 

In this work, we experimentally investigate the ability of dispersion managed (DM) designs to overcome the SC generation performances of fixed width (FW) and single tapered (ST) nanophotonic waveguides. Three different dispersion maps (DM, ST and FW) were optimally designed by numerical simulations of the nonlinear pulse propagation. We experimentally demonstrate octave-spanning supercontinua in the three configurations, and achieve maximum bandwidth from the edge of the silicon transmission window, $\sim$ 1.17$\mu$m, up to 2.76 $\mu$m by pumping at 2260 nm. Interestingly, the dynamics of the spectral broadening on the blue side of the spectrum in the ST and DM waveguides is different from that encountered in the FW one. This work, along with the recent achievements on integrated mode-locked lasers \cite{Zhou_2015,Wang_2017}, will have applications in power efficient, silicon-based, chip scale frequency comb generation and frequency synthesis.          

\section{\label{sec:level1} Experiment}
\subsection*{A. Method} 
The experimental setup for SC generation is shown in Figure 1. The frequency comb seed source consists of a Kerr-lens mode-locked Cr:ZnSe femtosecond laser (MODEL:CLPF-2200-10-70-0.8, IPG). The output wavelength of the laser can be tuned from 2130 nm to 2260 nm, close to the two-photon absorption (TPA) band-edge of the silicon at 2200 nm. The femtosecond laser provides a maximum average power of 1.2 W with 70 fs (full-width-at-half-maximum) pulses at 82 MHz repetition rate. A combination of a half-wave plate and a polarizer is used as a variable power attenuator and sets the polarization to excite the quasi transverse electric (TE) mode of the waveguide.

\begin{figure}
\centering
\includegraphics[width=8.6cm]{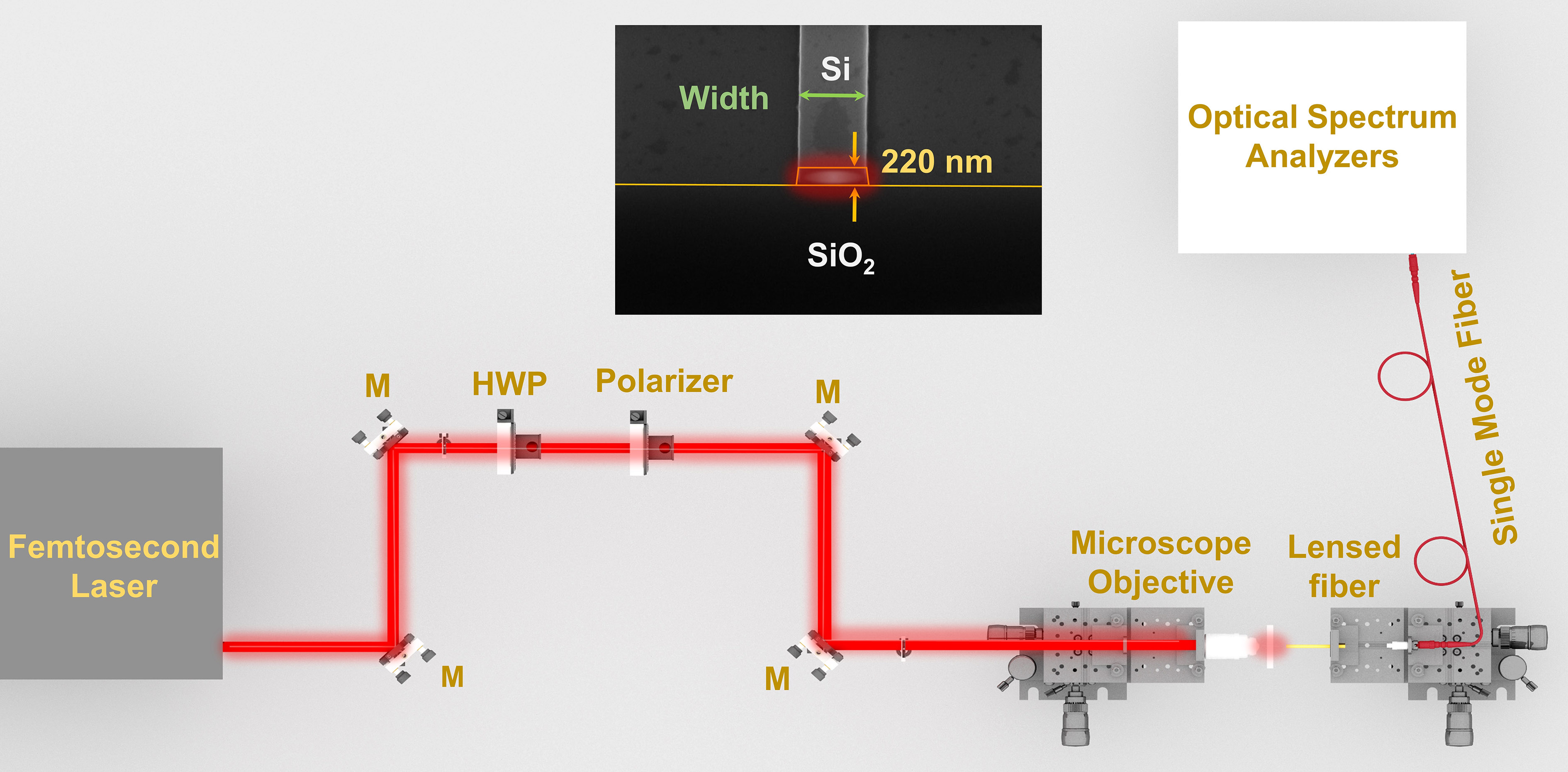}
\caption{Experimental setup for studying the SC generation in the nanophotonic waveguides (HWP, half wave plate; M, mirror). The inset shows the scanning electron microscope image of the cross-section of the silicon waveguide.}
\label{fig:1}
\end{figure}

The light is coupled into the waveguide by means of a microscope objective ($\times$60, NA = 0.65) and is collected by a silica based lensed fiber (working distance: 14 $\mu$m, NA = 0.4). This asymmetric coupling scheme ensures a close to transform limited input pulse and a large output collection efficiency. The lensed fiber to waveguide coupling efficiency is estimated at 8.3 dB at the pump wavelength from measurements at low power with symmetric in- and out-couplings with lensed fibers. A coupling efficiency of 24.5 dB through the microscope objective is obtained from linear transmission experiments. As the collected power does not vary significantly between the waveguides, the in- and out-coupling efficiencies are assumed constant for the different experiments. The output spectra are recorded using two optical spectrum analysers Yokogawa-AQ6370D (800-1700 nm) and AQ6376 (1500-3400 nm) to cover the entire supercontinuum bandwidth.

 \begin{figure*}
\centering
	\includegraphics[scale=1]{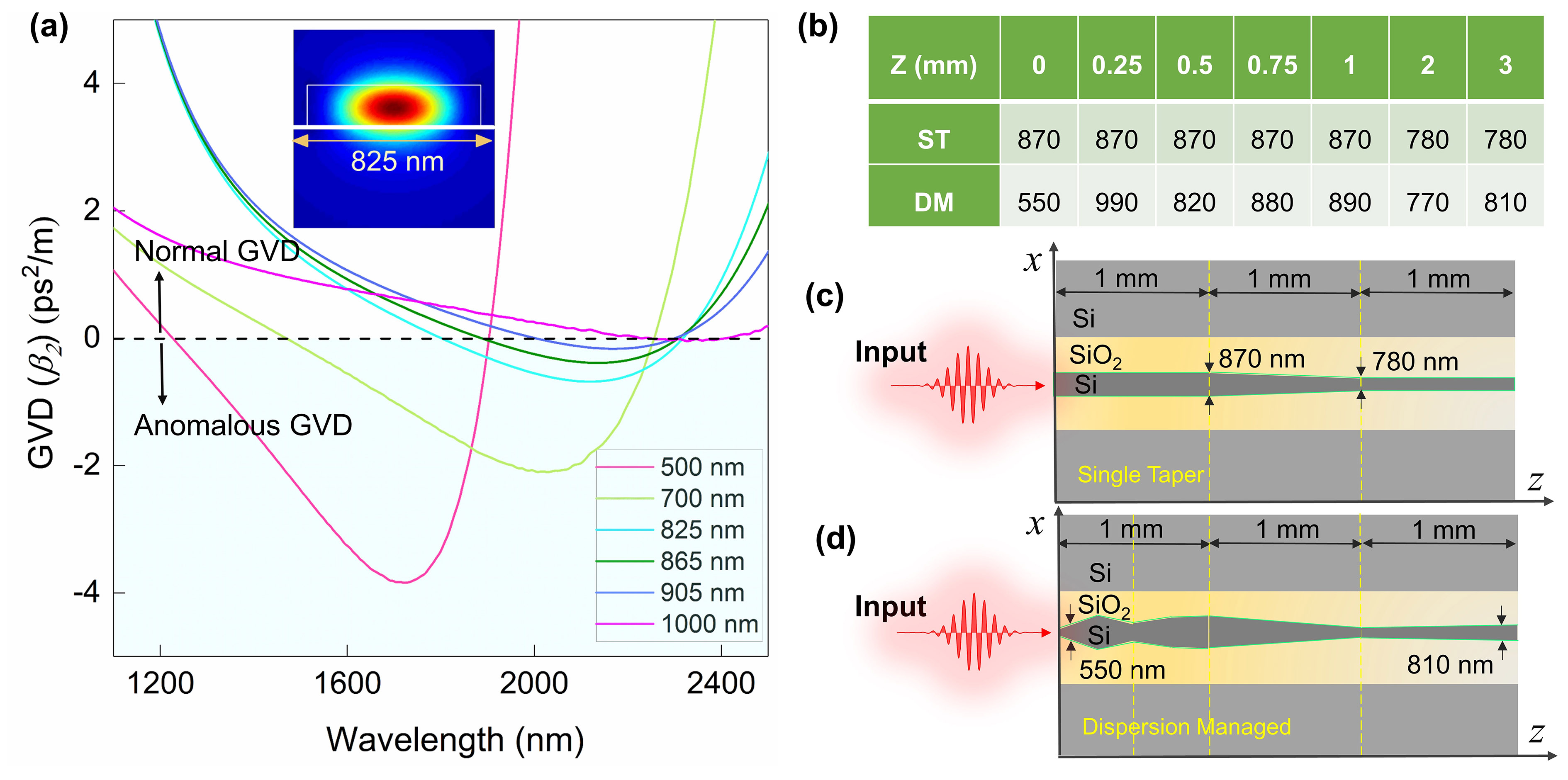}
	\caption{(a) Computed group-velocity dispersion of the fundamental quasi TE-mode of a 220-nm thick SOI waveguide with fixed width of 500, 700, 825, 865, 905 and 1000 nm. The inset shows the fundamental transverse mode in the 825 nm-wide waveguide at 2200 nm.  Right panel: (b) Dispersion map of the single tapered (ST) and of the dispersion managed (DM) waveguides obtained by the optimization algorithm. The table gives the nominal width at the adjustment points in nanometers as a function of their position (z) along the waveguide. The schematic geometry of the ST and DM structures are pictured in (c) and (d). }
	\label{fig:2}
\end{figure*}
 The air-clad SOI waveguides, fabricated in a CMOS pilot line, consists in a 220 nm-thick silicon layer on top of a 2 $\mu$m buried oxide layer. The inset in Figure 1 shows a scanning electron microscopy image of the cross-section of the silicon wire. For SOI nanophotonic waveguides, the normal dispersion of bulk silicon can be compensated by the geometric waveguide dispersion. In 220 nm-thick structures, the group velocity dispersion (GVD) depends strongly on the waveguide width and a net anomalous dispersion can be reached in the wavelength range ~ [1500 - 2300] nm. This indicates the possibility of dispersion management by modifying the silicon core geometry, similarly to fibers \cite{Zhang_2009,Arteaga_2014}. However, as shown in Figure 2 (a), while the first zero dispersion wavelength is highly dependent on the width, the second one, close to 2300 nm, much less so. Pumping around 2200 nm, the dispersion map is thus expected to mainly affect the blue side of the broadened spectra. The waveguides considered in this work have either a fixed or a variable width in the range [500 – 1000] nm, and are all 3 mm-long. The dispersion maps were designed following the same methodology as in our previous work [59]. For the single tapered (ST) waveguide, we impose the taper section to be located between 1 and 2 mm. The dispersion managed (DM) structure consists of several taper sections with the adjustment points (i.e. the positions where the slope changes) located at 0, 0.25, 0.5, 0.75, 1, 2, and 3 mm. The somewhat arbitrary choice of the number and location of those points is suggested by the typical dynamics encountered in fixed width waveguides in which the fission length is close to 1 mm, for the pulse duration and peak power considered. The optimum varying dispersion profile for each structure, i.e. the waveguide width at each adjustment points, was found by resorting to a genetic algorithm with the same fitness function as in \cite{Ciret_2017}, computed in the range [1200-2800] nm. This fitness function makes a trade-off between a broad and a flat output spectrum. 
 
  
The numerical simulations of the SC generation were performed by the integration of the generalized nonlinear Schrödinger equation \cite{Lau_2014}, taking into account the full dispersion curve computed with a commercial software (Lumerical, inc.), the optical Kerr and the Raman effects, the three-photon absorption (3PA), the free-carrier dispersion and absorption, the avalanche ionization and the shock term($\tau_{shock} = 1/\omega_0$). The nonlinear coefficients for the reference waveguide width of 800 nm are $\gamma = 220\,$W$^{-1}$m$^{-1}$ and $\alpha=0.022\,$cm$^{3}$GW$^{-2}$ at the pump wavelength \cite{Bristow_2007,Pearl_2008}. These coefficients are scaled with the effective mode area at other waveguide widths. For the design optimization, we considered a 70\,fs, 32\,W peak power hyperbolic secant input pulse centred at 2200 nm. The dispersion maps are given in Figure 2. For the fixed width waveguide, the simulation predicts an optimal width of 865 nm.  

\subsection*{B. Results} 
We start by generating supercontinua in waveguides with nominal fixed widths of 825\,nm, 865\,nm and 905\,nm. This will serve as benchmarks to compare with supercontinuum generation (SCG) in varying dispersion waveguides, as well as for calibrating the dispersion curves. Experimentally, the broadest spectra are obtained at a pump wavelength of 2260\,nm in the FW waveguide, but also in the DM and ST ones. The little shift from the design wavelength could be explained by small differences between the computed and the actual dispersion. This wavelength of 2260\,nm will therefore serve as the input pump wavelength from now on. Measured and simulated output spectra are plotted in Figures 3 (a), (b) and (d). 

 \begin{figure*}
	\includegraphics[scale=1]{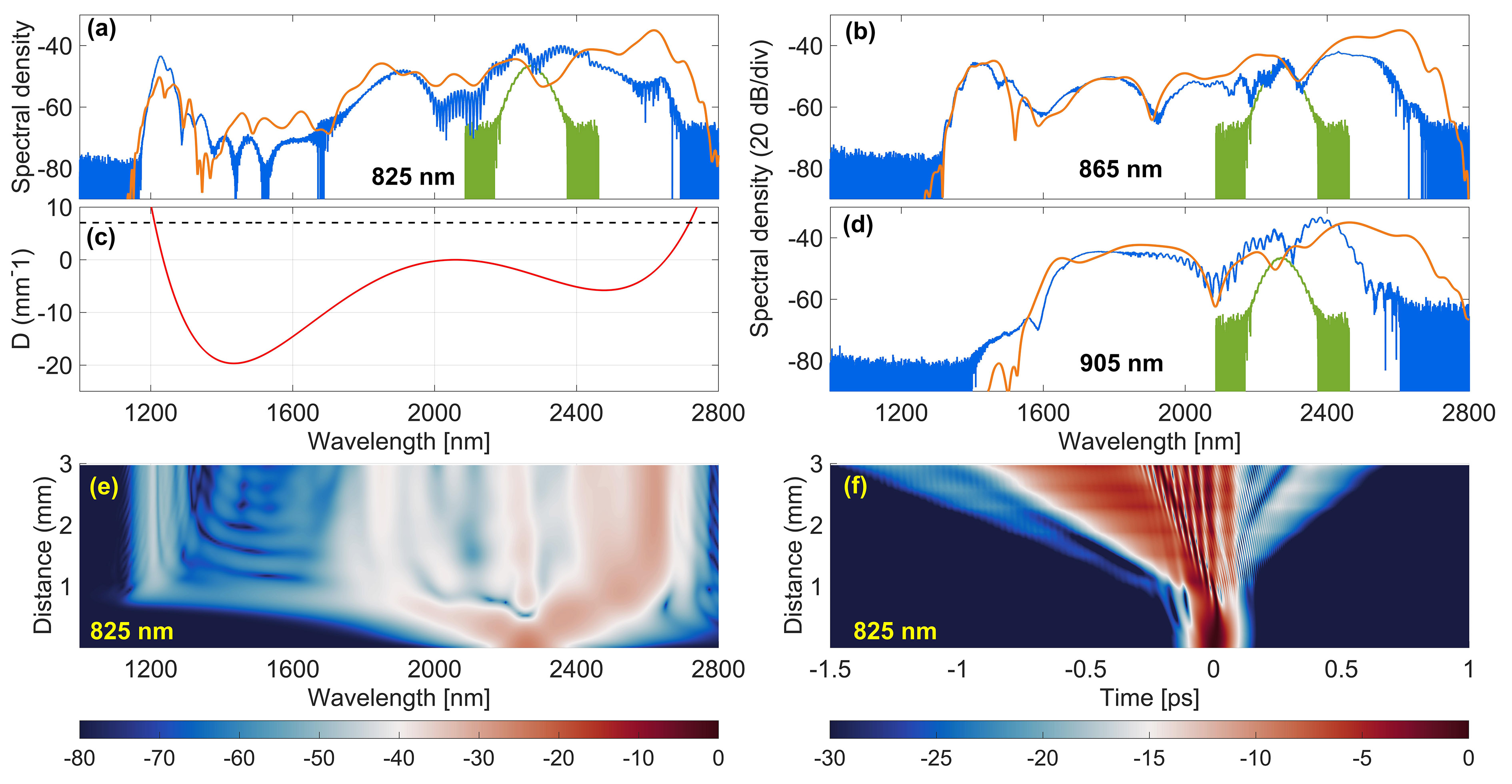}
	\caption{Experimental (blue) and simulated (orange) output spectra in the waveguides of nominal width (a) 825 nm, (b) 865 nm and (d) 905 nm for a pump pulse at 2260 nm (green). The widths used to compute the dispersion properties differ from the nominal ones. They are 830, 880 and 935 nm, respectively. The experimental average free space input power is 90 mW. }
	\label{fig:3}
\end{figure*}
In the simulations the input peak powers are 54, 30 and 40\,W, respectively. Figure (c) shows the wavenumber profile $D(\omega)$ of the 825 nm FW waveguide as used in the simulation, and plotted for a 2059 nm wavelength (see main text for further explanations). Resonant dispersive waves are emitted at wavelengths satisfying the phase matching condition Eq.(1), i.e. at the crossing points with the horizontal dashed line. Figures (e) and (f) show the evolution of the simulated spectral and temporal pulse profiles along the 825 nm FW silicon waveguide. 
  
In the 825\,nm FW structure, a distinct dispersive wave can be seen in the normal dispersion region on the blue side of the pump wavelength \cite{Ciret_2018}. This dispersive wave remains connected to the central part of the spectrum in the 865\,nm waveguide, but can barely be seen in the largest one, in agreement with the corresponding dispersion curves (see Fig. 2 (a)). The broadest spectrum is obtained in the 825\,nm FW waveguide for which the resonant dispersive wave is emitted the farthest from the pump wavelength. This results in a spectrum extending from 1168\,nm, close to the silicon band-edge, up to 2695\,nm. The numerical simulations of the SCG from a hyperbolic secant input pulse with 70\,fs duration agree well with the experiments, except beyond 2.6\,$\mu$m where a larger spectral density is predicted. This discrepancy is encountered in all waveguides, when the light is collected with the lensed fiber but also with a NIR microscope objective. It is thus attributed to higher losses than expected for large wavelengths at which the fundamental TE mode is mainly located around the lateral waveguide edges. To take into account the difference between the computed and the actual dispersion properties, we have considered the waveguide width as a free parameter. The widths are set to 830, 880 and 925\,nm in the simulations for the nominal widths of 825, 865 and 905\,nm, respectively. These values will later be used as a guideline to adjust the width profiles of the ST and the DM waveguides in the simulations. We can finally notice that the position of the dispersive wave generated at short wavelengths in the 825\,nm waveguide coincides very well with the phase matching condition \cite{Lau_2014} (see also Figure 3):
\begin{equation}
D(\omega_{DW})=\beta(\omega_{DW})-\beta(\omega_p)-(\omega_{DW}-\omega_p)/\upsilon_{gp} \approx 1/2P_p\gamma,
\end{equation}
where $\omega_{DW,p}$ are the frequency of the resonant dispersive wave and of the pump, $\upsilon_{gp}$ and $P_p$ are the group velocity and the peak power of the pump. The main peak of the temporally focused pulse at the fission point is here considered as the pump and corresponds to a 64 \,W peak power pulse at 2059\,nm. This $D(\omega)$ curve also predicts the emission of a second resonant dispersive wave at about 2650\,nm, which could well explain the small peak on the right edge of the experimental spectrum.We then consider the propagation in the single tapered and in the dispersion managed waveguides. In both cases, the spectra are more than one octave wide and extend from 1170\,nm to 2680\,nm in the ST waveguide, and from 1190\,nm to 2680\,nm in the DM one [see Figures 4(a) and (b)]. 

 \begin{figure}
	\includegraphics[scale=1]{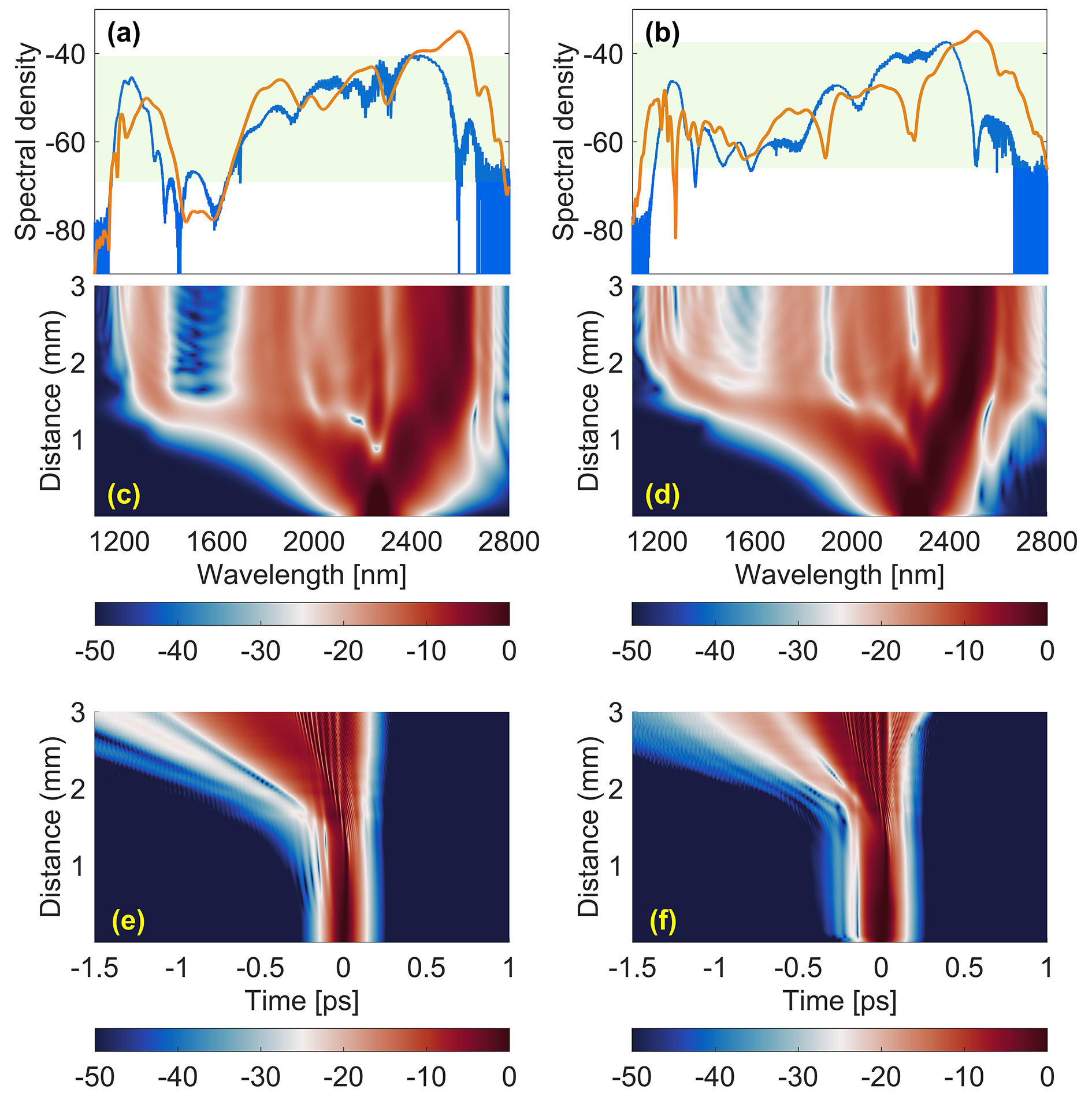}
	\caption{Supercontinuum generation in the single tapered (left column) and the dispersion managed (right column) waveguides. Figures (a) and (b) show the recorded (blue) and simulated (orange) spectra. The shaded area highlights a 30\,dB dynamic range. The waveguide widths at the adjustment points in the simulations are 885 and 780\,nm for the ST waveguide, and 555, 1010, 825, 895, 905, 775 and 815\,nm in the DM structure. Figure (c, d) and (e, f) show the corresponding evolution of the pulse in the spectral and the temporal domain. The average input power measured in free space is 70\,mW in both cases. In the simulations the input peak power is set to 30\,W for the ST waveguide and to 40\,W for the DM waveguide.}
	\label{fig:4}
\end{figure}

Looking at the output spectra, we can see that the SC generated in the DM waveguide is clearly flatter than in the ST and in the 825\,nm FW ones at similar input power, with at least 10\,dB more spectral density in the [1330-1700]\,nm wavelength range. Compared to the FW structure, the main difference in the dynamics of the SCG in the ST and DM structures can be made out in the numerical simulations of the spectral and temporal evolutions of the pulse profile. On the one hand, we can readily see the emission of the high frequency resonant dispersive wave in the 825\,nm FW waveguide in the temporal domain (with a positive delay), as it travels slower than the main pulse [see Figure 3 (f)]. On the other hand, this dispersive wave cannot be discerned in the ST one and it emerges only beyond 2.5\,mm in the DM structure [see figures 4 (e) and (f)]. This indicates that following its emission, this dispersive wave could interact with the main pulse, thanks to the axially varying dispersion properties. This is confirmed in the spectral domain. The dispersive wave, initially emitted around 1400\,nm, is frequency shifted toward 1200 nm, between $z\approx1.5$\,mm and $z\approx2$\,mm, i.e. in a decreasing taper section of the ST and of the DM waveguides. Since this frequency shift is more pronounced in the DM one, we will only focus on the dynamics in this structure. 
  
 \begin{figure}
	\includegraphics[scale=0.55]{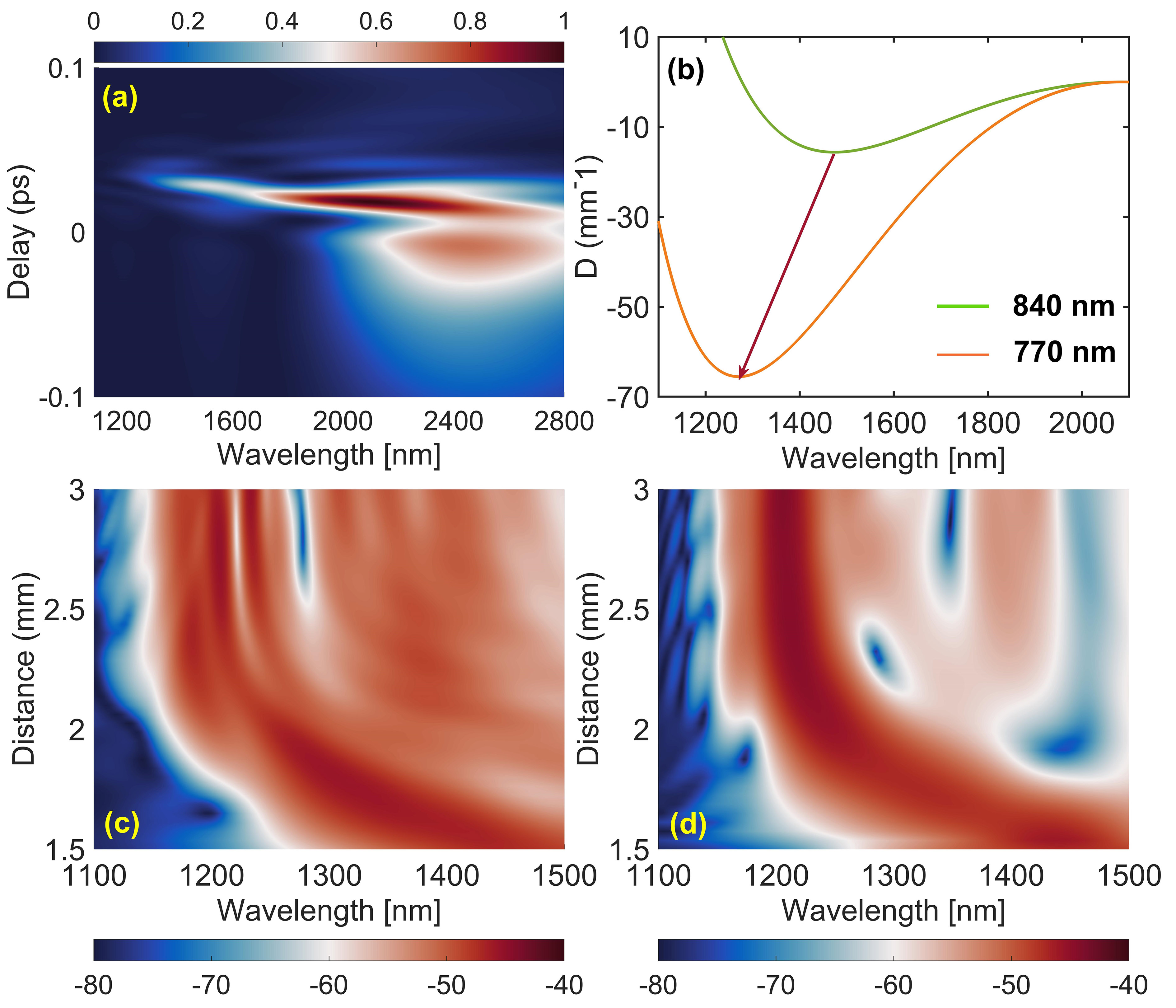}
	\caption{(a) Spectrogram of the pulse in the DM waveguide at $z=1.5$\,mm, starting from a 40\,W peak power, 70\,fs pulse at 2260\,nm. (b) Dispersion properties (wavenumber $D$) for a reference wavelength of 2085\,nm in 840\,nm and 770\,nm wide waveguides corresponding to the width at $z=1.5$\,mm and $z=2$\,mm in the DM structure. The arrow schematically shows the evolution of the group velocity matched wavelength in the taper section between those two locations. Numerical simulation of evolution of the spectrum in the DM waveguide between $z=1.5$\,mm and $3\,$mm: (c) same as in Fig. 4(d), and (d) when a 55\,W, 12\,fs pump pulse at 2085\,nm, together with a 5.5\,W, 20\,fs seed pulse at 1445\,nm, and with a 11\,fs relative delay is considered at $z=1.5$\,mm. In this latter case, the spectral shift of the seed pulse is similar to the spectral evolution seen in Fig.5 (c), confirming the trapping mechanism of the dispersive wave encountered in the DM waveguide. }
	\label{fig:5}
\end{figure}

In order to get a better insight into the interactions involved, we can first look at the spectrogram of the pulse at $z=1.5$\,mm, where the dispersive wave just emerges in the spectrum. The spectrogram plotted in Figure 5 (a) 
   \begin{figure*}
	\includegraphics[scale=1]{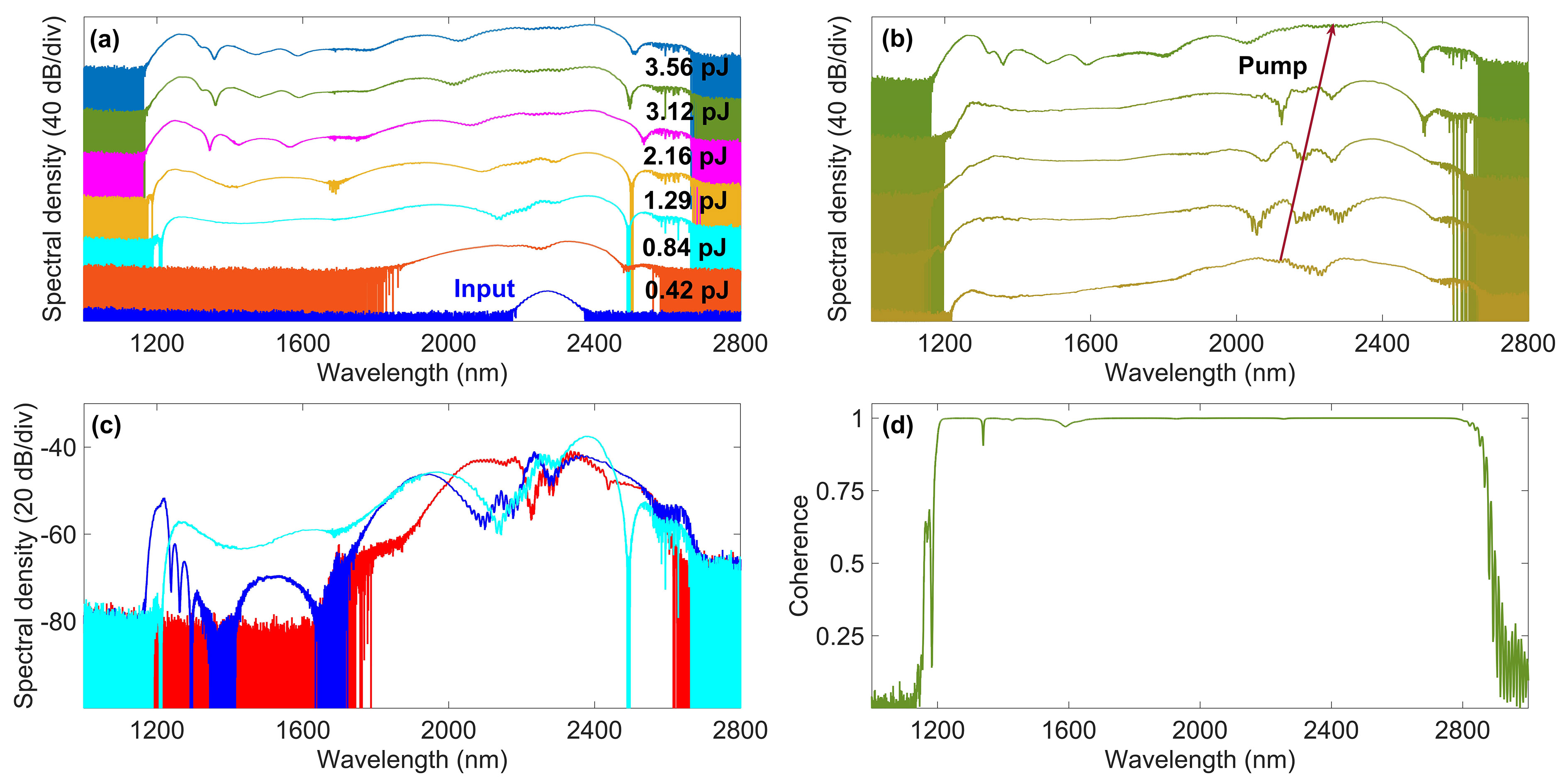}
	\caption{(a) Experimental output spectra for input pump pulse energies between 0.04\,pJ and 3.56\,pJ in the DM waveguide (bottom to top). (b) Measured output spectra at an input free space power of 70\,mW (3.12\,pJ coupled pulse energy) when the pump is tuned from 2140\,nm to 2260\,nm, as indicated by the red arrow. (c) Spectra recorded at the output of the DM (cyan), ST (blue) and 825\,nm-FW waveguides (red) for a free space average incident power of 19\,mW (0.84\,pJ coupled energy). (d) Computed degree of first order coherence of the SCG generated in the DM waveguide at 47\,W input peak power (or equivalently 3.7\,pJ pulse energy).}
	\label{fig:6}
\end{figure*}
%
shows that the focused pulse consists in a main short pulse, a blue detuned dispersive wave on the trailing edge of the main pulse and, finally, a wave component centred on 2.5\,$\mu$m in the normal dispersion region and travelling faster than the main pulse (see also Supplemental Material). By numerically filtering the main pulse and the blue detuned dispersive wave, we obtain a 55\,W peak power, 12\,fs pulse, centred on 2085\,nm and delayed by\,23 fs, and a smaller 5.5\,W, 20\,fs pulse, delayed by 34\,fs and centred on\,1445 nm. As seen in Figure 5 (d), the simulation of the propagation of two hyperbolic secant short pulses, with these parameters and an arbitrary relative phase, starting at $z = 1.5$\,mm in the DM waveguide nicely reproduces the main nonlinear dynamics occurring beyond that location, that is the continuous blue shift of the dispersive wave from 1445\,nm to 1200\,nm. This blue shift can be understood by looking at the dispersion curves $D$ plotted in Figure 5 (b). At $z = 1.5$\,mm the 1445\,nm wave is actually group velocity matched with the main pulse (same slope). It is thus up-shifted by cross-phase modulation with the trailing edge of the main pulse, and slowed down due to the dispersion. However, as the two pulses propagate in the decreasing taper, the up-shifted wave remains group velocity matched with the main pulse, leading to its continuous blue shift.  This non resonant process is known as intra-pulse interaction \cite{Gorbach_2006} and refers to the trapping of a dispersive wave by a decelerating soliton either by Raman scattering \cite{Gorbach_2007} or in tapered waveguides \cite{Travers_2009}. It differs from inter-pulse interactions which can also explain the enhanced blue shifted component of supercontinua observed in tapered waveguides \cite{Bendahmane_2015,Wang_2015}. This latter mechanism results indeed in a discreet frequency shift due to interactions at event horizon as observed in fibers \cite{Philbin_2008} or in integrated waveguides \cite{Ciret_2016}, and not a continuous frequency shift. Beyond $z = 2$\,mm, the group velocity matched wavelength stops decreasing and the non-resonant radiation lags behind the trailing edge of the main pulse as seen in Figure 4 (f).  
Further, we measured for the DM waveguide the dependence of the SCG with the tuning of the input power and of the pump wavelength. For potential applications of on chip generated broad supercontinuum, it would be desirable that their characteristics be robust, especially due to manufacturing tolerance issues. Figure 6 (a) shows the output spectrum for various input powers corresponding to estimated input pulse energies ranging between 0.04 and 3.56\,pJ. At an energy as low as 0.84\,pJ, the output spectrum is already very broad and smooth, contrary to the one generated in the ST and in the 825\,nm FW waveguides [see in Figure 6 (c)], demonstrating the value of the DM design compared to the ST and FW ones. Increasing the input power does not significantly change the output spectrum, probably because the spectrum is limited on the blue side by the absorption of the bandgap and, for large wavelengths, by the losses encountered in the waveguide. Since the identified mechanism behind the SCG in the DM structure is not limited to the considered geometry, broader spectra could in principle be obtained in silicon waveguides with a larger height and thus better guidance properties beyond 2.8\,$\mu$m. We then checked the tolerance regarding the input pump wavelength on the generated SC. The corresponding spectra are displayed in Figure 6 (b) and show that the excellent properties of the supercontinuum are maintained on the full tuning range of our pump laser. All generated spectra at 70\,mW input free space power (3.12\,pJ coupled energy) are octave spanning and extend approximately from 1200\,nm to\,2670 nm, as the pump is tuned from 2140\,nm to 2260\,nm. 
   \begin{figure*}
	\includegraphics[scale=0.98]{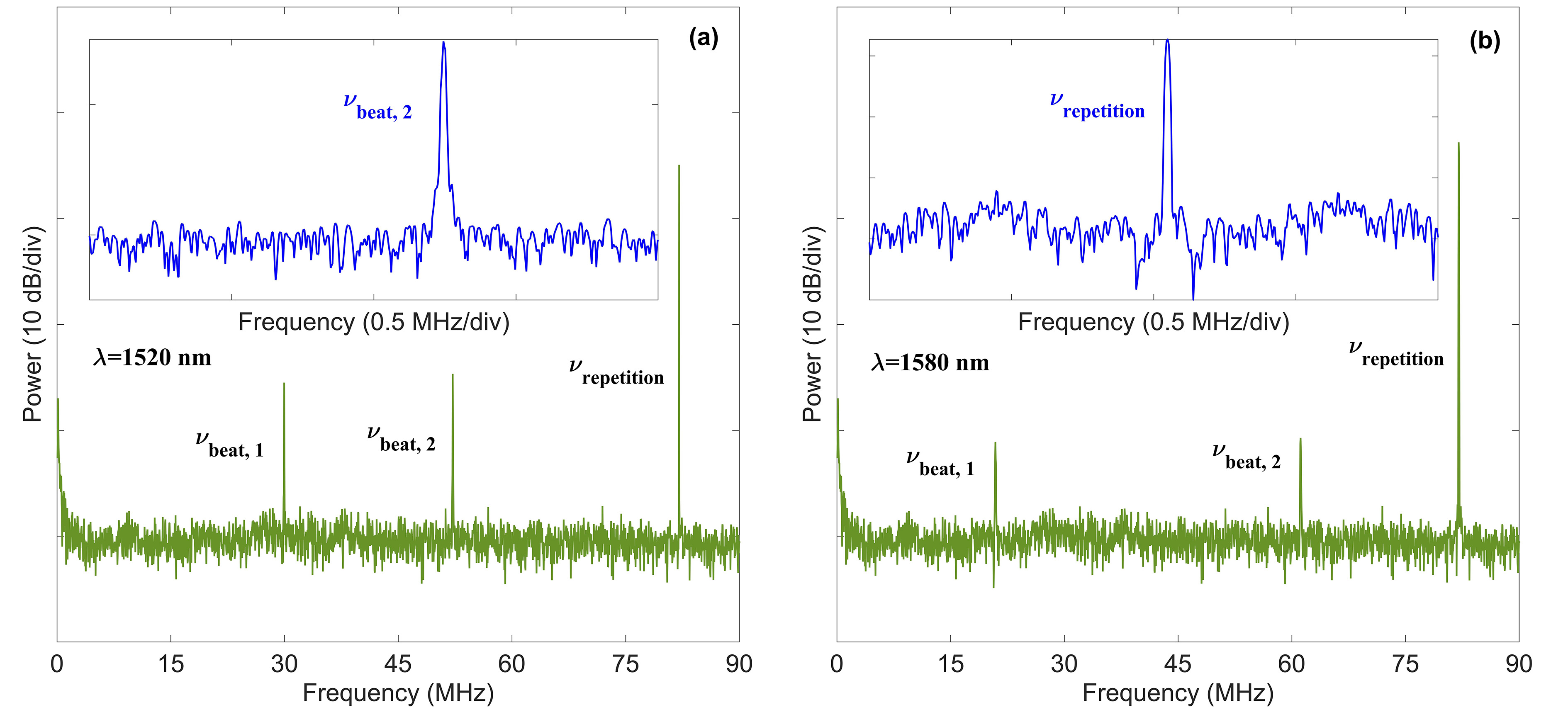}
	\caption{Beat note measured at the DM waveguide output with a CW wavelength at (a) 1520 nm and (b)1580 nm. The resolution bandwidth is 100 kHz. The insets show the spectrum of the beat note measured with a resolution of 10 kHz. }
	\label{fig:7}
\end{figure*}
 
Ideally, the supercontinuum would perform as a broad frequency comb that inherits the spectral coherence from the pump source. We have numerically investigated the output pulse coherence by adding to the input pulse a \textit{one-photon-per-mode} seed field with a randomized phase to each frequency discretization. The first-order coherence $|g_{12}|$ for 1000 independent pulse pairs was then calculated from the phase and amplitude variations of independent realizations \cite{Dudley_2002}. Figure 6 (d) shows that for a 47\,W peak power, 70\,fs input pulse, $|g_{12}|$ remains above 0.99 over the full bandwidth of the spectrally broadened pulse in the DM waveguide. A very high coherence is thus expected over the whole bandwidth of the supercontinuum. In order to confirm this, we have experimentally studied the phase coherence in the output spectrum generated in the DM structure. We measured the beat note between the SC and a tunable CW laser, an alternative approach to that involving the interference between independent realizations for assessing comb coherence properties \cite{Lu_2004,Ye_2010,Leo_2015}. The CW source consists in a narrowband laser (Santec TSL-770, $\textless$60\,kHz linewidth), tunable in the SCL telecom bands. This signal is then mixed with the supercontinuum by means of a 90/10 fiber coupler (bandwidth 1510-1590\,nm) and sent to a 125\,MHz bandwidth photodiode, sensitive in the range 900-1700\,nm. The RF signal from the photodiode is recorded by a 40\,GHz RF spectrum analyzer. The RF spectra measured with the CW laser at 1520\,nm and 1580\,nm are shown in Figure 7 (a) and (b) for the SC plotted in Figure 4 (b). Three isolated lines can be seen with the strongest one at 82\,MHz corresponding to the repetition rate of the pump laser. The two others are the heterodyne beat notes between the two closest lines of the supercontinuum and the CW laser. The spectrum acquired with a 10\,kHz resolution shows that the spectral width of the beating signal is about 50\,kHz and is thus limited by the linewidth of the CW laser. These results confirm the excellent coherence properties of the broadened frequency comb by supercontinuum generation, in the measured spectral band around 1550\,nm. We can finally notice that, in this spectral band, only the SC generated in the DM waveguide has a sufficient spectral density to perform beat note measurements [see Figs. 3(a) and 4 (a-b)]. The low power threshold to get broadband SC in DM structures and the high coherence of the resulting frequency comb is promising for robust chip-scale frequency comb sources, and contribute to the development of high-precision photonic devices for time and frequency applications.

\section{\label{sec:level1} Conclusion}

In this work we studied the ability of dispersion varying waveguides to achieve low power broadband coherent supercontinuum. The design of the best dispersion map to achieve broadband SC was based on numerical simulations of the pulse propagation and an optimization algorithm. Fixed width and single tapered structures were also considered. We experimentally demonstrated octave spanning supercontinuum generation from sub 100\,fs input pulse at 2260\,nm, covering the full transmission window of our waveguides. We showed that the SC generated in the dispersion-managed (DM) waveguide is as wide as with the single tapered and fixed width ones but is flatter, and requires less input power. From numerical simulations we infer that, in the DM waveguide, the dispersive wave on the blue side of the spectrum is first emitted close to the  pump wavelength. It is then trapped and continuously blue-shifted by the main compressed pulse as they both propagate in a taper section of the DM waveguide. This mechanism enables the efficient generation of the dispersive wave, as it is emitted closer to the pump, and then the spreading of its energy farther away by the trapping mechanism. It thus requires less pump power than in dispersion-engineered fixed width waveguides to achieve similar spectral broadening. We stress that it is the optimization of the dispersion profile, on the basis of the input pulse properties, that allows the dispersive wave, once emitted, to be continuously group-velocity matched with the pump, and thus trapped and blue-shifted. In addition, the output spectrum generated in our DM waveguide is robust against variations in input power (from ~20\,W to ~80\,W peak) or pump wavelength in the range [2140-2260]\,nm, and shows excellent coherence properties despite the encountered complex dynamics. Together with other demonstrations of silicon-based SC generation, this work establishes silicon as an important platform for SC generation. It demonstrates the benefit of dispersion managed structures over regular fixed width dispersion engineered waveguides and establishes them as promising scalable platforms for many applications, such as next-generation of photonic frequency synthesizers, high precision optical clocks, and spectroscopy devices.

\section*{\label{sec:level5} Acknowledgments}
This work was supported by the Fonds de la Recherche Scientifique - FNRS under grant $n^o$ PDR.T.1084.15. F.L acknowledges the support of the Fonds de la Recherche Scientifique (F.R.S.-FNRS, Belgium). B.K. and F.L acknowledge funding from the European Research Council (ERC) under the European Union’s Horizon 2020 research and innovation programme (grant agreement Nos 759483 $\&$ 757800).

\bibliography{DM_PRApp}

\end{document}